\documentclass[aps,prb,showpacs,groupedaddress]{revtex4}
\usepackage{graphics}
 
\usepackage{dcolumn}
\begin{document}

\title
{Stability of vortex state of Fulde-Ferrell-Larkin-Ovchinnikov type in magnetic fields perpendicular to superconducting layers}

\author{Ryusuke Ikeda}

\affiliation{%
Department of Physics, Graduate School of Science, Kyoto University, Kyoto 606-8502, Japan}

\date{}

\begin{abstract}
Motivated by recent NMR data suggesting the presence of a Fulde-Ferrell-Larkin-Ovchinnikov (FFLO) vortex lattice in magnetic fields  perpendicular to the superconducting plane in the heavy-fermion superconductor CeCoIn5, we examine here the stability of FFLO states in quasi two-dimensional (2D) superconductors under a perpendicular field. It is found that a 3D-like portion of Fermi surfaces peculiar to CeCoIn5 stabilizes a FFLO vortex state modulating along the field parallel to the c-axis and that the direction of modulation in a FFLO vortex lattice is related to the characters of the normal to FFLO and the Abrikosov to FFLO transitions.   
\end{abstract}

\maketitle

\section{Introduction}

Recently, a series of observations \cite{1,2,3} in the heavy fermion superconductor CeCoIn5 at low temperatures and in high fields seem to have clarified that the new high field phase in the parallel field configuration (${\bf H} \parallel  ab$), surrounded by the discontinuous $H_{c2}$-transition and a continuous transition from the Abrikosov vortex lattice, will be a Fulde-Ferrell-Larkin-Ovchinnikov (FFLO) vortex lattice. This FFLO state \cite{4} is expected to have a one-dimensional modulation of the gap function along\cite{5} ${\bf H}$ through the observation \cite{2} of a reduction of ultrasound velocity in the new phase, which can be attributed to an additional vortex tilt assisted by an elastic fluctuation of the nodal planes of the FFLO structure \cite{6}. The presence of the nodal planes seems to be also consistent with the NMR data \cite{3}. 
A more recent NMR measurement \cite{7} in perpendicular fields (${\bf H} \parallel c$), together with a subtle anomaly in heat capacity data \cite{1}, has shown features similar to the above-mentioned ones in ${\bf H} \parallel ab$. This similarity strongly suggests the presence of a FFLO vortex state with a modulation along the $c$-axis and the applied field. However, this fact may be unexpected, because the Fermi surface (FS) of CeCoIn5 is roughly cylindrical, and the FFLO modulation in the Pauli-limited case (i.e., with no vortices) is not expected to develop along a direction with open FS. 
We point out here that the presence of a small sheet of three-dimensional (3D) FS can support a FFLO modulation parallel to ${\bf H}$, which is stable with respect to a vortex tilt. Further, the modulation direction of a possible FFLO state is found to be reflected in the characters of the $H_{c2}$ 
and FFLO transitions. 

\section{Phase diagram and Fermi surface}

Our analysis is based on a microscopic derivation of a Ginzburg-Landau functional and is a straightforward extention \cite{6} of that in Ref.5, where a 2nd order FFLO to Abrikosov transition field decreasing upon cooling was first demonstrated. Typical examples of phase diagrams we obtain at this time are shown in Fig.1 where $h$ denotes the magnetic field normalized by the orbital limiting field. Except for the used forms of FS, the material parameters in Fig.1 (a) and (b) are the same as each other. Figure 1(a)  follows from the use of a purely cylindrical FS, in which the in-plane component $(V_F)_{ab}$ of the Fermi velocity is much larger than $(V_F)_c$ everywhere, and, as in the Pauli-limited case, the FFLO modulation tends to be formed along the ab-plane everywhere in the phase diagram. Due to the Landau-level (LL) structure of the pair-field in the presence of vortices, however, a two-fold symmetric structure due to the nodal planes parallel to ${\bf H}$  is not described in the lowest LL ($n=0$) but rather in the next LL ($n=1$). The nodal planes can be regarded as being accommodated as periodic stripes in the $n=1$ vortex lattice \cite{9}. That is, a FFLO state modulating not along ${\bf H}$ but in a perpendicular direction to ${\bf H}$ may occur at low enough temperature ( $t \equiv T/T_{c0} < 0.15$ ) in Fig.1(a). However, the normal to n=1 vortex lattice transition, occurring on the thick dashed line below $t < 0.15$,  is of 2nd order.  
\begin{figure}[t]
\scalebox{0.42}[0.45]{\includegraphics{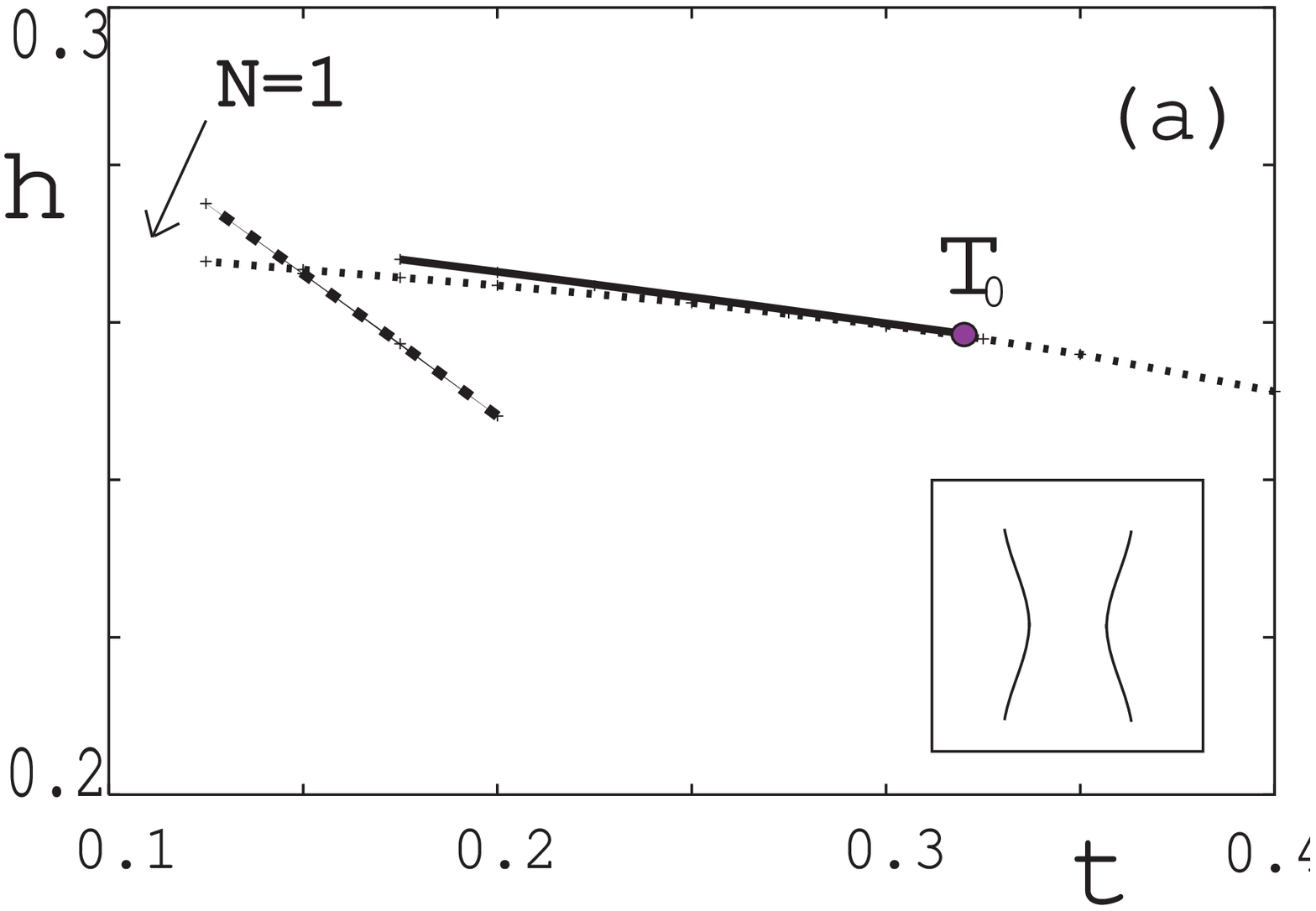}}
\scalebox{0.42}[0.45]{\includegraphics{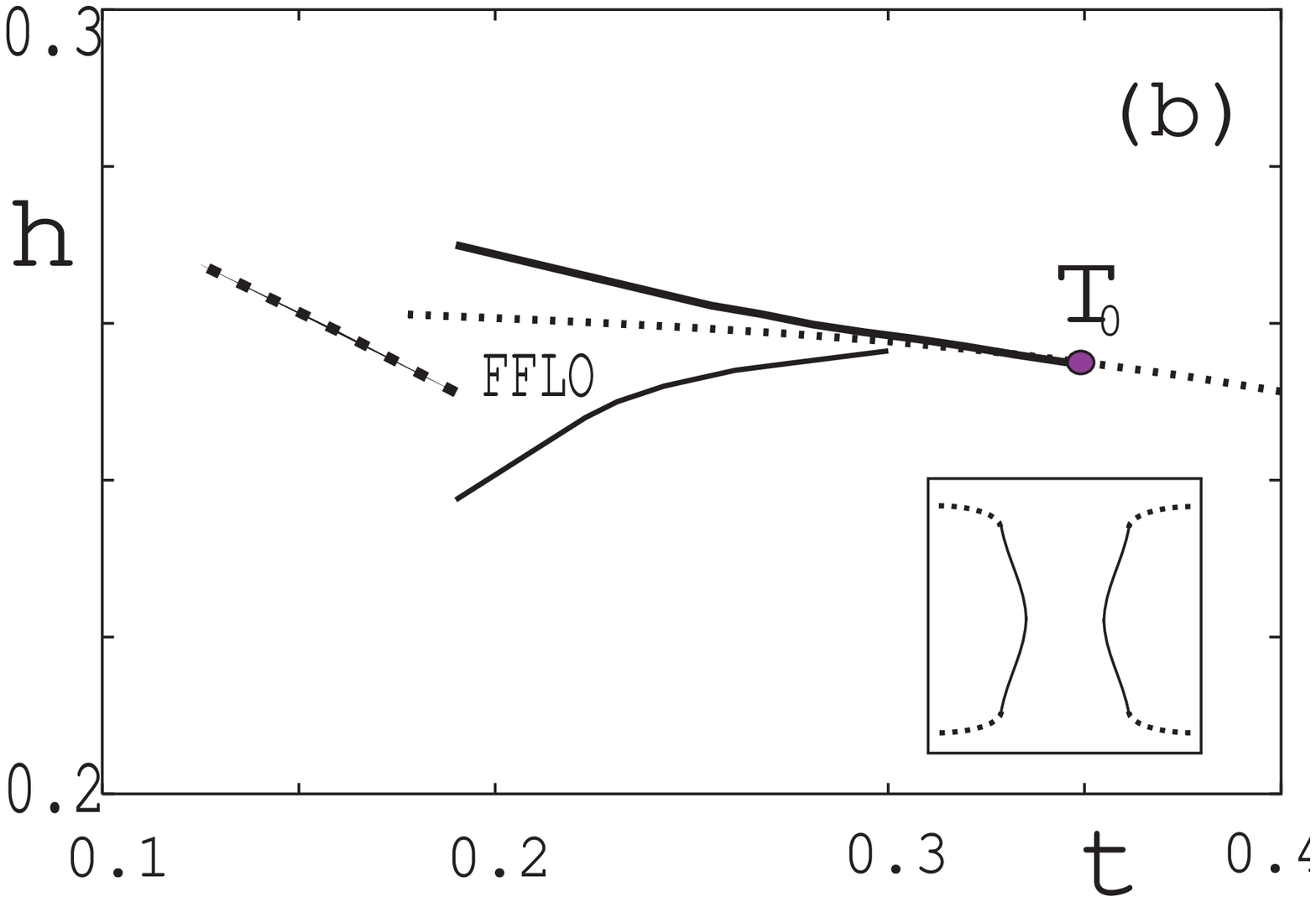}}
\caption{Examples of $h$ v.s. $t$ ($= T/T_{c0}$) phase diagrams in ${\bf H} \parallel c$ resulting (a) from the purely cylindrical FS and (b) from a cylindrical one with 3D-like portions (see the dotted lines in the Inset). Below $T_0$, the $H_{c2}$-transition occurs on each thick solid curve and is discontinuous, while each dotted curve is the 2nd order $H_{c2}$-transition curve and its extrapolation. The thin solid curve in (b) is the 2nd order transition line between the Abrikosov vortex lattice and the FFLO one with modulation $\parallel {\bf H}$. In both figures, the Maki parameter is 1.0.}
\label{fig: }\end{figure}
On the other hand, in Fig.1(b) where a cylindrical FS with small 3D-like branches  (the dashed lines in the Inset) 
was used, we have a 2nd order transition between the Abrikosov vortex lattice and a FFLO one modulating 
along ${\bf H}$ and with nodal planes perpendicular to ${\bf H}$ in much higher temperatures than the region dominated by the $n=1$ LL. This FS in Fig.1(b) simulates that expressed by the $\beta_1$ and $\beta_2$ bands \cite{8} of the FSs of CeCoIn5. As observed \cite{7} in CeCoIn5 , the transition from the normal to this FFLO state is discontinuous \cite{5}, while the FFLO to Abrikosov transition is of 2nd order. Notice, by comparing both the figures with each other, that the directions of modulation in FFLO states correlate with the characters of the transitions surrounding the FFLO state {\it when the vortices are present}. 

\section{Discussion}
 
Several remarks are needed to compare Fig.1(b) with the experimental phase diagram \cite{7}. First, although the FFLO state with nodal planes $\parallel {\bf H}$ inevitably appears at the lowest temperatures where the $H_{c2}$-transition is of 2nd order, a small amount of impurities play the role of shifting various temperature scales to lower temperatures \cite{6} so that the $n=1$ LL state is lost in real systems. Second, the experimental 2nd order Abrikosov to FFLO transition line is flat and insensitive \cite{7} to $t$ in contrast to that in Fig.1(b). However, we find \cite{6} that a flat FFLO transition curve is obtained by slightly reducing the 3D portion of FS or by incorporating the antiferromagnetic (AF) fluctuation because an AF fluctuation, weaker with increasing $h$, competes with the Pauli paramagnetic effect. Further, we note that the FFLO state modulating along ${\bf H}$ in ${\bf H} \parallel c$ is more stable than that in ${\bf H} \parallel ab$. In the latter, the nodal plane becomes softer with increasing $h$ or the material anisotropy, and a reduction \cite{2,6} of tilt modulus occurs in the FFLO state, while the tilt modulus in the FFLO state in the former rather increases upon cooling \cite{6}. Ultrasound experiments in ${\bf H} \parallel c$ are useful for checking the validity of the present theory.

The author thanks Y. Matsuda for informative discussions. 

{99} 

\end{document}